# Characterization of 3D-DDTC detectors on p-type substrates

Gian-Franco Dalla Betta, *Senior Member, IEEE*, Maurizio Boscardin, Luciano Bosisio, Giovanni Darbo, Paolo Gabos, Claudia Gemme, Michael Koehler, Alessandro La Rosa, Ulrich Parzefall, Heinz Pernegger, Claudio Piemonte, *Member, IEEE*, Marco Povoli, Irina Rachevskaia, Sabina Ronchin, Liv Wiik, Andrea Zoboli, *Student Member, IEEE*, Nicola Zorzi

*Abstract* – We report on the electrical and functional characterization of 3D Double-side, Double-Type-Column (3D-DDTC) detectors fabricated on p-type substrates. Results relevant to detectors in the diode, strip and pixel configurations are presented, and demonstrate a clear improvement in the charge collection performance compared to the first prototypes of these detectors.

## I. INTRODUCTION

SILICON 3D detectors consist of an array of columnar electrodes of both doping types, oriented perpendicularly to the wafer surface and penetrating entirely through the substrate [1]. This architecture offers a number of advantages with respect to the planar one, among them low depletion voltage and fast collection times, while keeping the active thickness unaltered. As a result, 3D detectors are intrinsically radiation-hard and are emerging as one of the most promising technologies for the innermost layers of tracking at the foreseen upgrades of the Large Hadron Collider [2],[3].

Since 2004, in the framework of a collaboration between FBK-irst and INFN, we have developed modified 3D architectures aimed at a simplification of the manufacturing technology with respect to the original Full-3D design. After introducing and carefully evaluating the 3D-STC (Single-Type-Column) approach [4], in the past three years we have been investigating 3D-DTTC (Double-side, Double-Type-Column) detectors [5], which involve columnar electrodes of both doping types etched from different wafer sides (junction columns from the front side and Ohmic columns from the back side), and stopping at a short distance (a few tens of μm) from the opposite surface. A similar approach is being pursued by CNM-IMB (Barcelona, Spain [6]).

At the 2008 NSS we have reported on the functional characterization of the first batch of this type of detectors, which was made on 300-μm thick, n-type substrates [7]. Although the column depths were not optimized (the overlap between n- and p-type columns was just 60 μm), encouraging results have been obtained from strip detectors connected to the 40MHz ATLAS SCT Endcap electronics (ABCD3TA chip): using a $Sr^{90}$ source setup, the efficiency at 1 fC threshold was 94% (mainly due to loss from the hollow columns) and the maximum collected charge was 2.4 fC. After irradiation with 24 MeV protons at a fluence of $2\times10^{15}$ $cm^{-2}$, the collected charge was still 1.5 fC, limited by trapping and ballistic deficit, the latter being due to the presence of low field regions resulting from the non optimized column depths.

Since then, we have fabricated at FBK two batches of 3D-DDTC detectors on 200-μm thick, FZ p-type substrates. These detectors feature a larger overlap between columns of opposite type. Thus low field regions are reduced in size and good charge collection efficiency is expected even with fast read-out electronics.

In this paper, after recalling the main design and technological properties of 3D-DDTC structures, we report on selected results from the electrical and functional characterization of detectors in the pad, strip and pixel configuration.

## II. TECHNOLOGY AND SIMULATIONS

3D-DDTC detectors fabricated on p-type substrates have $n^+$ read-out columns arranged in the diode, strip or pixel configuration using surface connections made with $n^+$ diffusion and/or metal strips. On the contrary, Ohmic columns ($p^+$) are all connected together by surface doping and metallization on the back side. All columns have a nominal diameter of 10 μm and are not filled with poly-Si. Surface isolation between $n^+$ columns on the front side is obtained by a combination of p-spray and p-stop implants. The schematic cross-section of a detector is shown in Fig. 1.

Manuscript received November 12, 2009. This work was supported in part by the Autonomous Province of Trento and in part by the Italian National Institute for Nuclear Physics (INFN), Projects TREDI (CSN5) and ATLAS (CSN1).

G.-F. Dalla Betta, P. Gabos, M. Povoli, and A. Zoboli are with INFN, Sezione di Padova (Gruppo Collegato di Trento), and with Dipartimento di Ingegneria e Scienza dell'Informazione, Università di Trento, Via Sommarive, 14, I-38123 Povo di Trento (TN), Italy (telephone: +39-0461-883904, e-mail: dallabe@disi.unitn.it).

M. Boscardin, C. Piemonte, S. Ronchin, and N. Zorzi are with Fondazione Bruno Kessler (FBK-irst), Centro per i Materiali e i Microsistemi, Via Sommarive, 18, I-38123 Povo di Trento (TN), Italy (telephone: +39-0461-314458, e-mail: boscardi@fbk.eu).

L. Bosisio and I. Rechavskaia are with INFN, Sezione di Trieste, and Dipartimento di Fisica, Università di Trieste, via A. Valerio, 2, I-34127 Trieste, Italy (telephone: +39-040-3756262, e-mail: bosisio@ts.infn.it).

M. Koehler, U. Parzefall, and L. Wiik are with the Institute of Physics, University of Freiburg, Hermann-Herder-Str. 3, 79104 Freiburg, Germany (telephone: +49-761-2035933, e-mail: Michael.Koehler@physik.uni-freiburg.de).

G. Darbo and C Gemme are with INFN, Sezione di Genova, Via Dodecaneso 33, 16146 Genova, Italy (telephone: +39-010-3536454, e-mail: Giovanni.Darbo@ge.infn.it).

A. La Rosa and H. Pernegger are with CERN – PH, CH-1211 Geneve 23, Switzerland (telephone: +41-22-76-75847 , e-mail: alessandro.larosa@cern.ch).

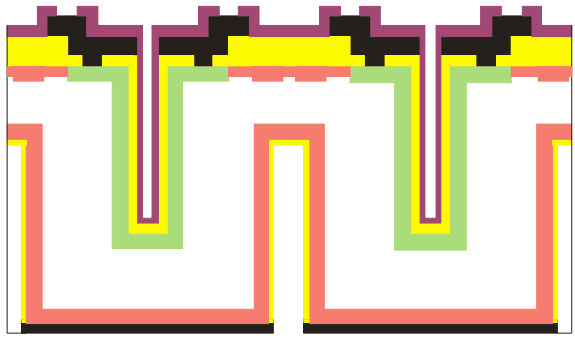

Fig. 1. Schematic cross-section of a 3D-DDTC detector fabricated on a p-type substrate (not to scale).

The fabrication technology is basically the same reported in [5] for p-on-n devices, apart from the substrate type, the inverted order of the column doping steps, and the additional ion implantations on the front side for p-spray/p-stop isolation. Deep Reactive Ion Etching (DRIE) is used for columnar electrodes. For the first batch of detectors (3D-DTC-2), this process step had to be performed as an external service at IBS (France), a fact which caused major delays in the fabrication and also set some constraints in the maximum etching depths. More recently, a new DRIE equipment (Adixen AMS200) has become available at FBK-irst. Using this equipment, a fabrication recycle (3D-DTC-2B) could be processed completely in-house.

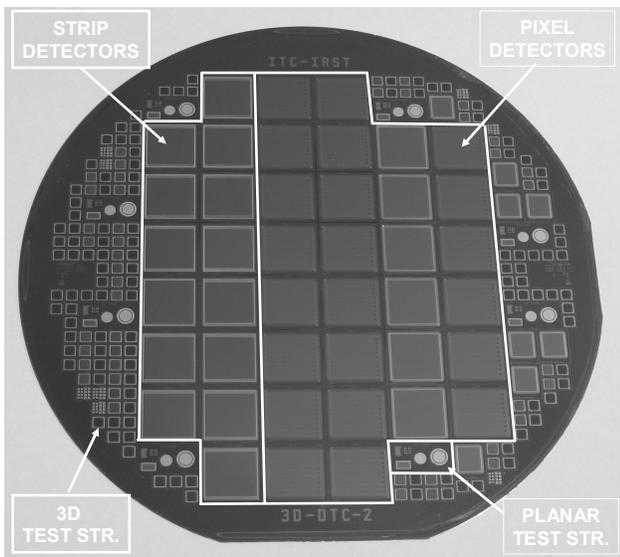

Fig. 2. Photograph of a processed wafer from the 3D-DTC-2B batch.

Figure 2 shows the photograph of a processed wafer, with blocks of different structures indicated. The wafer layout is mainly oriented to pixel detectors compatible with existing ATLAS and CMS single readout chips, but it still contains strip detectors and test structures (among them, planar and 3D diodes).

From Scanning Electron Microscopy (SEM) analysis and capacitance-voltage electrical measurements on test structures, the depth of read-out columns and Ohmic columns for batch 3D-DTC-2 was found to be about 100 μm and 190 μm, respectively [8]. For batch 3D-DTC-2B, the etching depth increased up to 170 μm for the read-out columns (but varied somewhat from wafer to wafer) and remained about 190 μm for the Ohmic columns, so that a larger column overlap was achieved. In those detectors having maximum column depth, the electric field configuration is comparable to that of Full-3D detectors, so that a very good charge collection efficiency is expected [5].

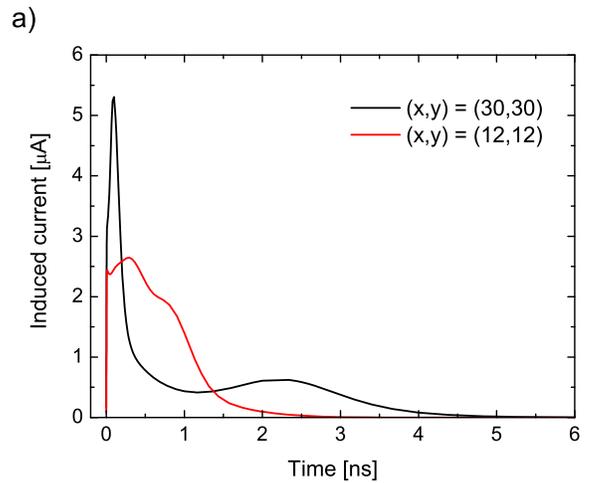

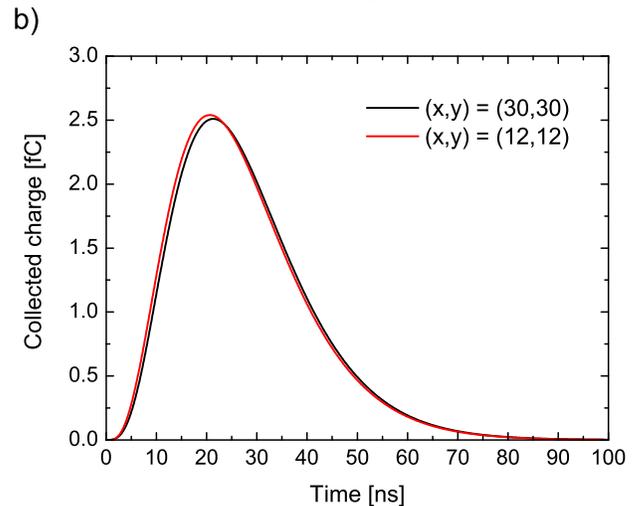

Fig. 3. Simulated transient signals in 3D-DDTC detectors biased at 15 V in response to a minimum ionizing particle hitting at t=0 perpendicularly to the surface in two different positions: (x,y)=(30,30), i.e., close to the junction column (which is at (40,40)), and (x,y)=(12,12), i.e., close to the Ohmic column (which is at (0,0)): a) current signals; b) equivalent charge signals at the output of a semi-gaussian CR-(RC)$^3$ shaper with 20 ns peaking time.

This is also confirmed by three-dimensional device simulations performed with Synopsys TCAD tools. As an example, Fig. 3a shows the simulated currents induced by a minimum ionizing particle impinging on the detector perpendicularly to the surface at t=0 in two different positions: (x,y)=(30,30), i.e., close to the read-out column (collecting electrons), and (x,y)=(12,12), i.e., close to the Ohmic column (collecting holes). The bias voltage is 15 V, i.e., just above full depletion voltage. In both cases, the current signals are just a few ns long, allowing for good charge collection efficiency also for fast read-out, as confirmed by Fig. 3b, where results from a post-processing of the current signals of Fig. 3a are shown, accounting for a semi-Gaussian CR-(RC)$^3$ filter which emulates a fast, ATLAS-SCT-like readout electronics with 20 ns peaking time.

## III. EXPERIMENTAL RESULTS

### A. Test Structures

Results from the electrical characterization of planar and 3D test structures indicate a good process quality. The leakage current is low, in the order of a few nA/cm$^2$, corresponding to less than 1 pA/column in 3D structures. Lateral depletion voltage between columns is as low as 3 – 4 V, whereas full depletion voltage is in the range from 10 to 15 V. The intrinsic breakdown voltage value (due to the p-spray) is close to 100 V. Due to some defects, a few detectors have lower breakdown voltage, but in all cases the breakdown voltage is at least one order of magnitude higher than the full depletion voltage. A comprehensive report on the results of the electrical characterization of test structures from the first batch can be found in [8]. Very similar results have been obtained from the second batch [9].

### B. Diode Detectors

In order to assess the detector signal dynamics, transient current measurements were performed at the University of Trento on 3D diodes using fast (~1 ns FWHM) laser pulses at different wavelengths. Diodes are arrays of 20×20 columnar electrodes with 80 μm pitch between columns of the same type (40√2 ≅ 56 μm pitch between columns of opposite type), with a total area of 2.56 mm$^2$. The laser spot is focused down to 20-30 μm onto the diode surface. Current signals were read out using a fast transimpedance amplifier featuring a gain of 100 V/A.

Figure 4 shows the integral of the current pulses as a function of time for three different laser wavelengths (the optical power was adjusted to yield approximately the same charge signal in all cases). The diode reverse bias is 30 V. As can be seen, the fastest response is observed for the 980 nm laser, due to the fact that its absorption length is about 100 μm, i.e., similar to the read-out column depth, so that most of the charge is generated in the high-field region where columns overlap. At 1060 nm, absorption length is much larger than the substrate thickness, so that the charge deposit can be considered quite uniform through the detector thickness. In this case, charge collection is slightly slower, because also the lower field regions are involved. At 850 nm wavelength, the absorption length is much lower (~17 μm), so that the top regions above the Ohmic columns are mainly involved, where the electric field is lower, this delaying the device response. In all cases, measured charge collection times are worse than predicted by simulations, a fact that is to be ascribed to the non optimized setup and limited amplifier bandwidth. Nevertheless, the response time is low enough to avoid ballistic deficit problems when read-out with fast, LHC-like electronics. Further details can be found in [10].

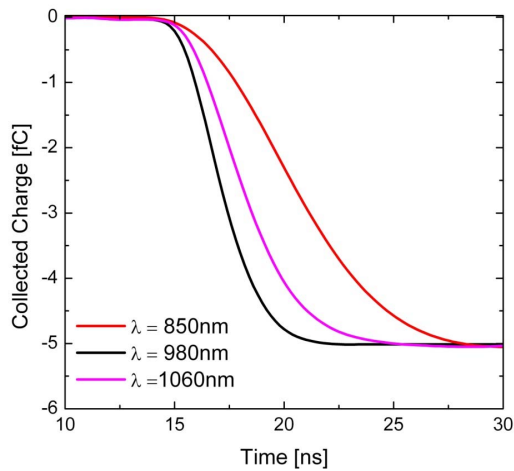

Fig. 4. Integral of current pulses vs time at different laser wavelength for 3D diodes biased at 30 V and read out with a fast transimpedance amplifier.

### C. Strip Detectors

Strip detectors feature 102 strips, each one containing 102 read-out columns, with 80 μm pitch between them. Also in this case the pitch between columns of opposite type is 40√2 ≅ 56 μm. The total strip length is 8.2 mm. Strips are AC coupled by means of integrated capacitors and biased by punch-through at both ends from a bias ring surrounding them. Each strip has four AC pads (two at each end) and one DC pad. It should be noticed that, due to the large punch-through gap between the strip ends and the bias ring (21.5 μm), which is determined by the combined p-spray/p-stop isolation, the voltage to be applied to the bias ring in order to reach lateral (20 V) and full depletion (52 V) in strip detectors are significantly larger than those measured on 3D diodes.

Strip detectors assemblies based on the ATLAS ABCD3T binary readout [11], with 20ns shaping time, and running at 40MHz, have been functionally tested at the University of Freiburg. Here we report noise and position resolved laser signal measurement. Due to the adopted binary read-out, both noise and signal values are determined by performing scans of the discriminator threshold, yielding the so called s-curves, which represent the normalized hit efficiency versus threshold value. The median amplitude of the analog signal is normally assumed to be the threshold value corresponding to the 50% point in the s-curve, whereas the noise is determined from the width of the s-curve. Figure 5 shows the average equivalent noise charge (ENC) as a function of the reverse voltage for both bonded and unbonded channels of the read-out chip. For bonded channels, noise is strongly dependent on the strip capacitance. The latter was also measured directly on a few strips of the detector using a probe station. Fig. 6 shows the interstrip and backplane components of the capacitance and their sum. The overall agreement between the ENC and the total capacitance curves is good. In particular, the ENC difference between bonded and unbonded channels at voltages larger than 40 V corresponds to a strip capacitance of 7.5 pF, a value which is very close to that measured electrically.

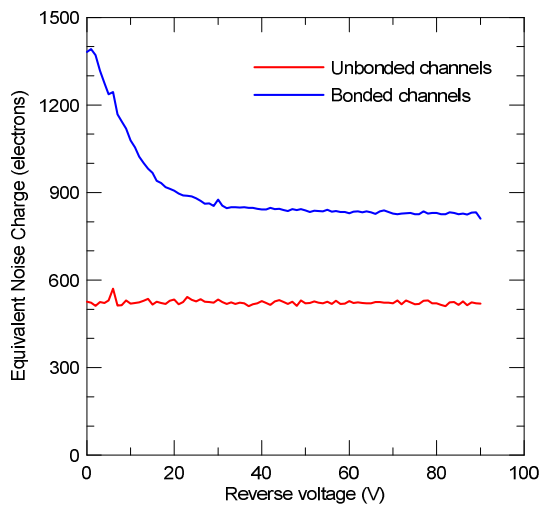

Fig. 5. Average Equivalent Noise Charge as a function of reverse voltage for unbonded and bonded channels of the read-out chip.

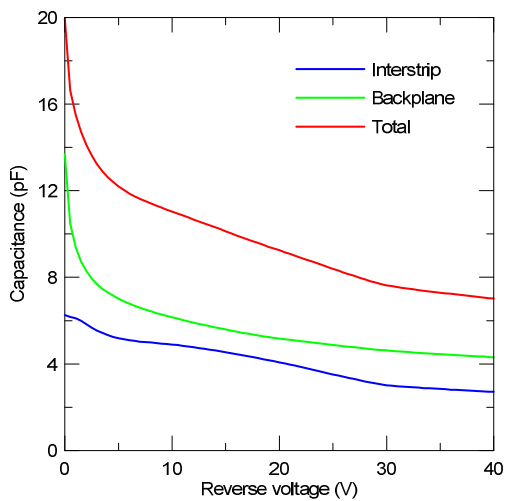

Fig. 6. Strip capacitance as a function of reverse voltage, showing interstrip and backplane contributions and total value. Data refer to electrical measurements performed at 8 MHz signal frequency.

In order to investigate the uniformity in the detector response, which was one of the main problems in 3D-STC detectors and, to a lower extent, was also affecting the first prototypes of 3D-DDTC detectors [7], position resolved laser measurements were performed on strip detectors. The test setup is based on an infrared (peak wavelength 982 nm) laser with a 2ns pulse width. Laser focusing on the detector surface to a spot size of about 4 μm is achieved by using a microscope. Remote-controlled motorized stages allow the sensor module to be moved in the x–y plane with μm-accuracy. A 100 μm × 100 μm square region including four read-out columns and one Ohmic column has been scanned (see Fig. 7). It should be noticed that the structure of 3D-DDTC detector lends itself to a non uniform signal response because of columnar electrodes. Columns cannot provide any signal, because they are empty, but since they are not passing-through the entire substrate thickness, signals can actually be observed from regions in between the column tips and the opposite surface. Additionally, surface layers can play a role in these tests: in fact, light is reflected by metal layers, causing zero-signal regions, and some differences in the light transmission are also expected from those regions featuring a different thickness of the passivation layers (e.g., the p-stop regions).

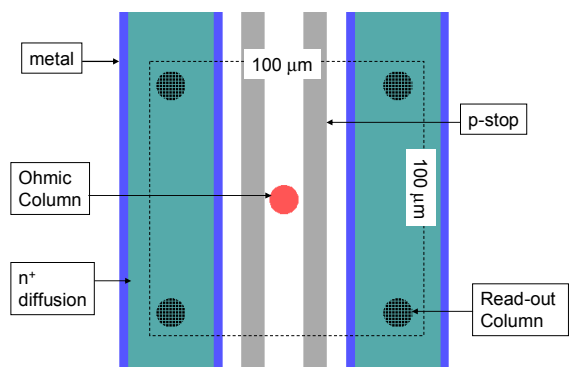

Fig. 7 Layout detail of strip detectors, showing the region of interest for position resolved laser measurements (dashed square).

Figure 8 shows high-resolution scans of the region of interest indicated in Fig. 7, performed at different bias voltages. The left strip signal is shown. At all voltages, zero signal from the light reflecting metal regions can be observed, as well as signals from regions beneath the junction column tips. As reverse voltage is increased, wider portions of the sensor show high signals due to the spreading of depletion region towards the Ohmic column in the centre of the cell. At 20 V, lateral depletion is reached and signal starts to saturate almost everywhere. This is due to the fact that light absorption length is comparable to the column overlap depth. Comparing results at 20 V and 80 V, it can be seen that the only regions showing increased signal are those beneath the columns, where depletion regions further proceed vertically toward the backplane. As expected, lower signal is observed in the centre of the cell due to the Ohmic column, and higher signal is observed along the two p-stops, because of the higher transmission efficiency of the photons at the interface, resulting from a lower oxide thickness. Further details about position resolved laser measurements as well as results from charge collection efficiency tests with a β source setup on strip detectors can be found in [12].

### D. ATLAS Pixel Detectors

ATLAS pixel sensors compatible with the present read-out chip FEI3 [13] have been fabricated. Single chips (i.e., arrays of 18 × 160 pixels) have been fabricated, featuring different column configurations, with two (2E), three (3E), and four (4E) read-out columns per pixel. The layout of these pixels is shown in Fig. 9. The pixel size is 400 μm × 50 μm.

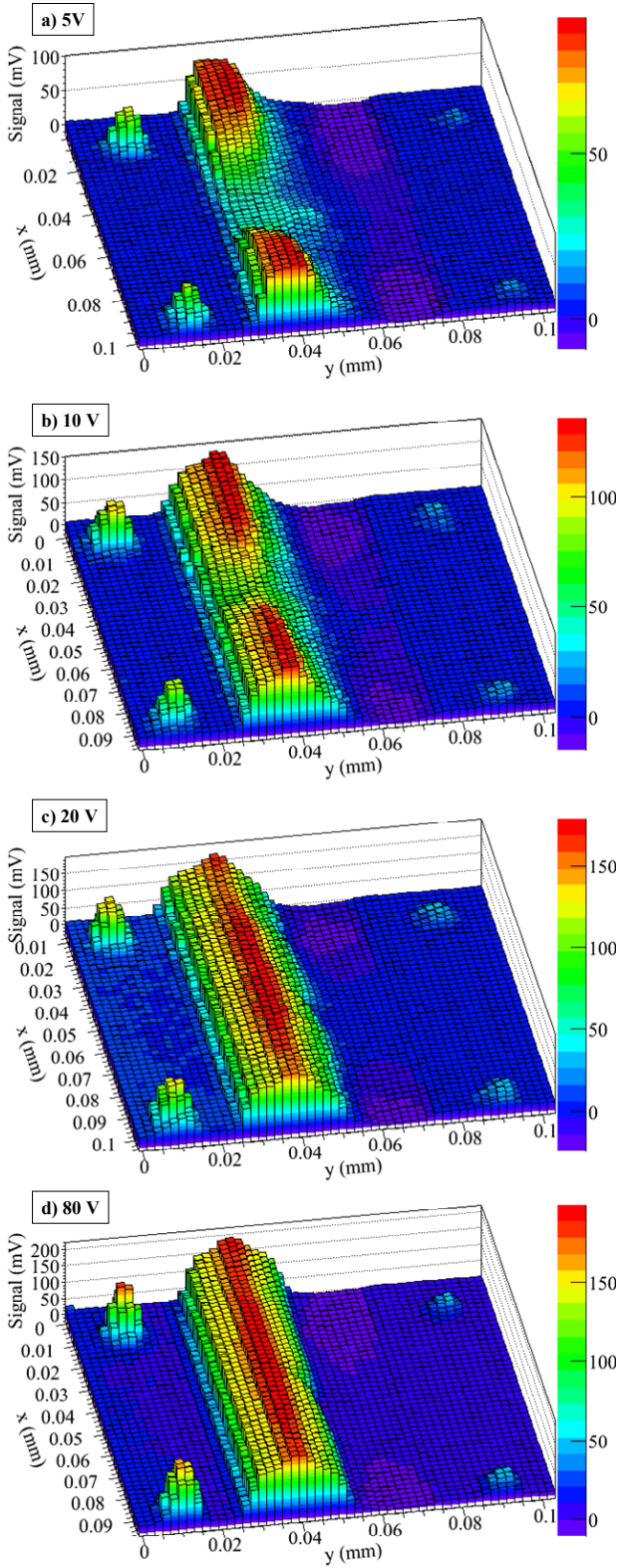

Fig. 8. Maps of the signals measured on the left strip at different bias voltages: a) 5 V, b) 10 V, c) 20 V and d) 80 V. The region of interest is shown in Fig. 7. Laser scan resolution is 2 μm both in x and y directions.

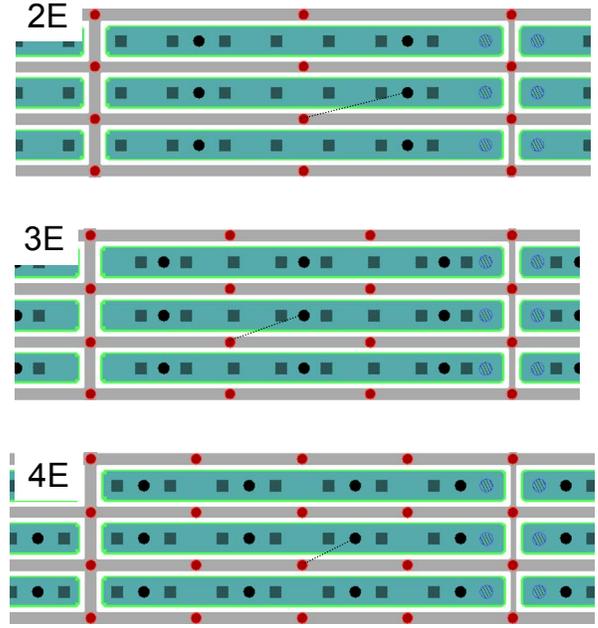

Fig. 9. Layout of ATLAS pixels featuring different column configurations (read-out columns are black circles, Ohmic columns are red circles). The distance between read-out columns and Ohmic columns is 103 μm, 71.2 μm, and 56 μm for 2E, 3E, and 4E, respectively.

Some pixel sensors from both fabrication batches have been connected via bump bonding to the ATLAS FEI3 read-out chip at SELEX SI [14]. Figure 10 shows a snapshot of a single-chip card.

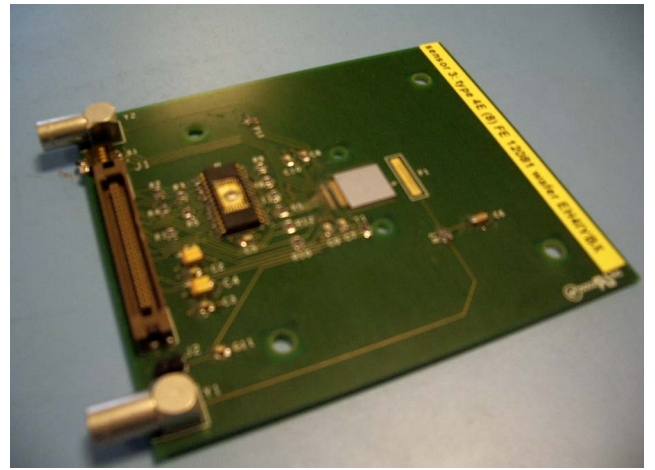

Fig. 10. Photograph of a single chip card.

In the FEI3 chip, the in-pixel front-end includes a charge amplifier and a discriminator with programmable threshold. The digital readout provides information on the hit pixel address, the hit time stamp and the digitized amplitude. Signal amplitude is expressed using the Time-over-Threshold (ToT), measuring the duration of the discriminator output pulse in units of the 40 MHz clock period. The ToT depends on the deposited charge, the discriminator threshold and the feedback current [13].

Laboratory tests on pixel sensors have been carried out at INFN Genova and CERN using the TurboDAQ setup [15], based on the NI LabWindows development suite. This setup is the same used for automated electrical tests of ATLAS Pixel Detector Modules during the production phase. The performed tests include leakage currents, threshold and noise, and response to radioactive sources.

The leakage current of pixel sensors from the first batch was found to be lower than 1 µA, corresponding to a pixel current in the range from 100 pA to 200 pA. If compared to values measured on diode and strip detectors, these leakage current values are sizeably larger, although still reasonable. The current degradation is believed to come from the bump bonding process. Apart from two defective sensors, showing early breakdown, all sensors can be operated up to at least 70 V, as expected from measurements of test structures. Sensors from the second batch have shown similar leakage current values, but could not be operated at voltages larger than 12 V due to early breakdown problems [9]. In the following, results from functional tests will be reported only for sensors belonging to the first batch.

Threshold and noise measurements have been performed on each pixel, inducing a signal by means of on-chip charge injectors, and recording the number of collected hits versus injected charge. Similarly to what previously described for strip sensors, threshold and noise are determined by fitting an s-curve to the threshold scan results of each pixel. Threshold dispersion is reduced by repeating the scans and adjusting a DAC-parameter individually for each channel [13], [16].

The ToT signal is calibrated using the on-chip injection circuits. The standard tuning aims at reaching 60 ToT units for a charge of 20,000 electrons. Choosing a standard threshold of 3,200 electrons, the corresponding charge is in the range from 250 to 350 electrons per ToT unit. Further details can be found in [9].

The noise as a function of the reverse voltage in three sensors with different column configurations is shown in Fig. 11. In non irradiated sensors, the main contribution to noise is known to come from the capacitance. As a matter of fact, different noise values are measured in sensors with different layouts, in agreement with the basic expectation that a larger number of columns per pixels corresponds to a larger capacitance, also due to the smaller pitch between columns of opposite doping types (see Fig. 9).

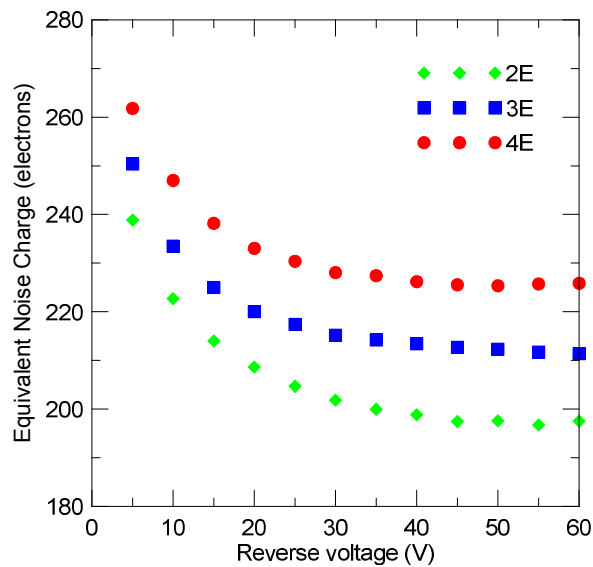

Fig. 11. Equivalent Noise Charge as a function of reverse voltage in three sensors featuring different column configurations.

In all cases, noise is found to decrease with reverse voltage, because the capacitance is also decreasing with voltage, until saturation can be observed in the voltage range from 30 V to 60 V, depending on the layout. It should be stressed that, although lateral depletion between columns occurs at low voltage (from about 4 V in 4E to about 13 V in 2E), the pixel capacitance keeps on decreasing in a much wider voltage interval. This effect could be directly observed by measuring the capacitance in a strip-like test structure reproducing the same layout of ATLAS pixels detectors.

As an example, Fig. 12 shows the capacitance as a function of voltage as measured in the test structure featuring the 4E pixel layout. As can be seen, the capacitance is not yet saturated at 30 V, especially because of the backplane contribution. We attribute this behavior to the fact that, due to the non optimal column depth configuration (the read-out column depth is about 100 µm, to be compared to the 200 µm substrate thickness), full depletion at the bottom of the wafer close to the Ohmic columns can be achieved only at a reverse voltage much larger than the lateral depletion voltage, as predicted also by TCAD simulations [9].

When scaling the capacitance values measured on test structures to the size of real pixels, the resulting values are found to be in very good agreement with the measured noise data (as an example, see the inset in Fig. 12). The saturation values of the pixel capacitance are 250 fF, 310 fF, and 370 fF for 2E, 3E, and 4E, respectively.

Finally, it should also be noticed that noise values reported in Fig. 11 are lower than those reported in [3] for full 3D sensors from the Stanford Nanofabrication Facility coupled to the FEI3 chip, mainly because of a lower capacitance resulting from a lower column overlap, and only slightly higher than those reported for planar pixel sensors (160 e, [17]).

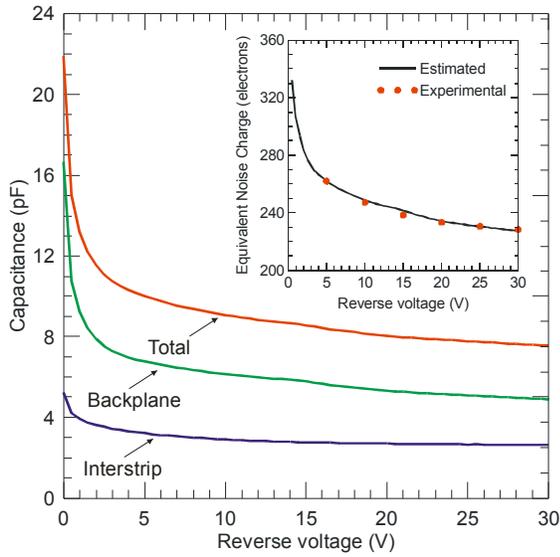

Fig. 12 Capacitance as a function of reverse voltage in a strip-like test structure featuring 4E layout with 82 read-out columns. Interstrip and backplane components, and total capacitance are shown. Data refer to electrical measurements performed at 8 MHz signal frequency. The Equivalent Noise Charge vs reverse voltage curve of the 4E pixel sensor estimated on the basis of the pixel capacitance is shown in the inset and compared to experimental values.

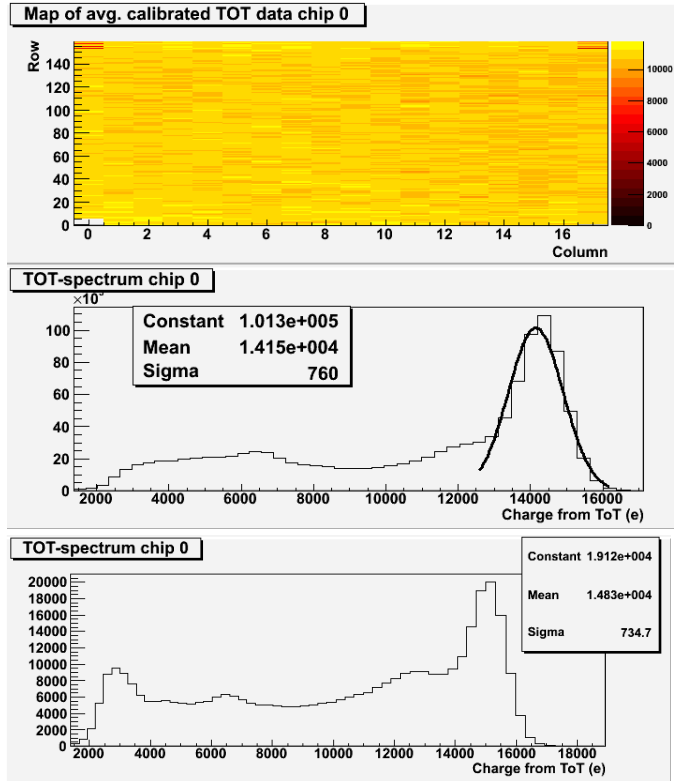

Fig. 13. Map of the charge signal from all pixels in a 4E sensor biased at 35 V and exposed to an $Am^{241}$ source (top); corresponding spectrum (centre); spectrum acquired from a planar pixel detector reverse biased at 200 V. Spectra have been obtained using the self-trigger capabilities of the FEI3 chip and summing signals from all pixels without any clustering. The reported statistical data refer to a Gaussian fit of the main peak.

Sensor calibration has been carried out performing tests with radioactive X-ray sources. As an example, Fig. 13 shows a map of the measured signals and the corresponding spectrum for a 4E sensor reverse biased at 35 V and exposed to an $Am^{241}$ source. The mean value of the collected charge for the main peak in the distribution is about 14,100 electrons. In silicon, a signal of about 16,500 electrons would be expected from the absorption of a 60 keV photon. The discrepancy between the two values is to be attributed to the uncertainty in the calibration process, which was estimated to be in the order of 15%, as confirmed by the reference spectrum acquired from a planar ATLAS pixel sensor using the same setup, also shown in Fig. 13.

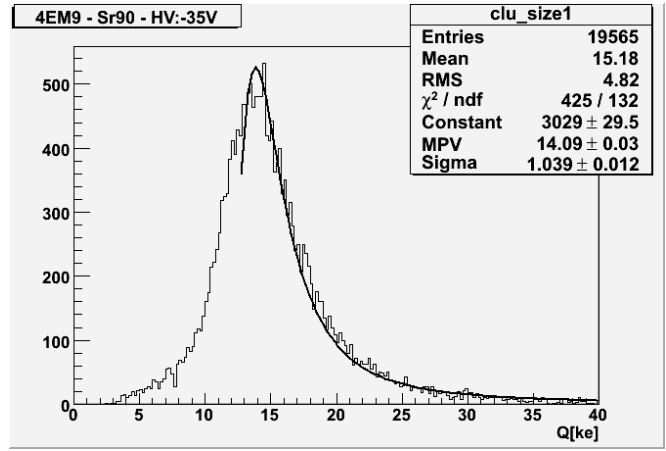

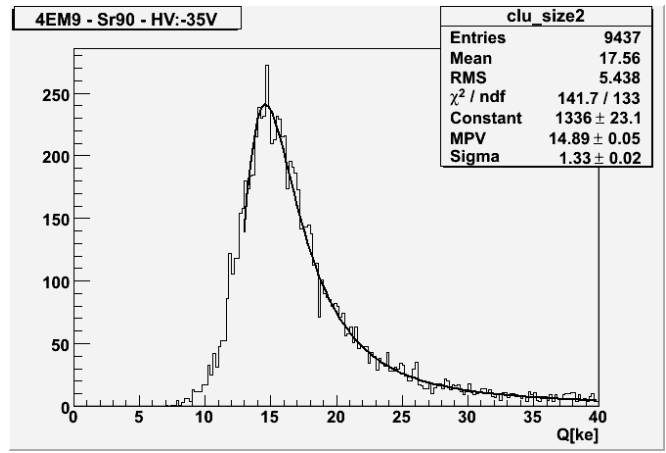

Fig. 14. Pulse height distributions for cluster size 1 (top) and cluster size 2 (bottom) in a 4E sensor biased at 35 V in response to a $Sr^{90}$ source. The reported data refer to a Landau fit to the charge spectrum.

Charge collection tests have also been performed using a $Sr^{90}$ β source. Figure 14 shows the pulse height distributions in a 4E sensor biased at 35 V for the two cases of cluster size 1 (top) and cluster size 2 (bottom). Charge values have been fitted with Landau functions, also shown in the figures. The most probable value (MPV) of the charge is ~14,100 electrons for cluster size 1 and increases up to ~14,900 electrons for cluster size 2. The latter spectrum also suppresses the low charge tail observed in the cluster size 1 distribution and which is likely due to charge sharing.

Similar charge distributions and MPV have been measured in 2E and 3E sensors [9]. This is not surprising, since the different column configurations are not expected to yield different charge collection behavior in non irradiated devices.

The reference planar pixel sensor has also been tested and the measured MPV values are about 17,200 (18,500) electrons for cluster size 1 (cluster size 2). Keeping in mind the different thickness of the planar sensor (250 μm), the MPV values measured in 3D-DDTC sensors are found to be properly scaling with the sensor thickness.

## IV. CONCLUSION

We have reported on the characteristics of 3D-DDTC detectors made on p-type substrates. Owing to a large overlap between columns (still to be optimized) and to electron read-out, these detectors have shown a sizeably enhanced performance with respect to the first prototypes presented at NSS 2008, in good agreement with TCAD simulation predictions. Ballistic deficit is no longer observed even using fast, LHC-like read-out electronics and the signal response is quite uniform, apart from the columns themselves. The capacitance values are higher than for planar sensors: this results into a relatively high noise for strip detectors, but only marginally affects noise for pixel sensors. Full charge collection efficiency in response to radioactive sources has been proved, and should now be validated after irradiation.

## ACKNOWLEDGMENT

We would like to thank: G. Gariano, A. Rovani ed E. Ruscino (INFN Genova, Italy), F. Rivero (University of Torino, Italy), and J.-W. Tsung (University of Bonn, Germany), for their precious help in system assembly and measurements; R. Beccherle (INFN Genova, Italy) for designing bump bonding mask; S. Di Gioia (Selex SI, Italy) for the bump bonding process.

## REFERENCES

[1] S. I. Parker, C. J. Kenney, J. Segal, "3D - A proposed new architecture for solid-state silicon detectors", *Nuclear Instrum. Methods A*, vol. 395, pp. 328-343, 1997
[2] C. Da Via, S.I. Parker, G. Darbo, "Development, Testing and Industrialization of Full-3D Active-Edge and Modified-3D Silicon Radiation Pixel Sensors with Extreme Radiation Hardness Results, Plans", ATLAS Upgrade Document, available http://cern.ch/atlas-highlumi-3dsensor
[3] C. Da Viá, et al., "3D active edge silicon sensors with different electrode configurations: Radiation hardness and noise performance", *Nuclear Instrum. Methods A*, vol. A604, pp. 505-511, 2009.
[4] C. Piemonte, et al., "Development of 3D detectors featuring columnar electrodes of the same doping type", *Nuclear Instrum. Methods A*, vol. 541, pp. 441-448, 2005.
[5] A. Zoboli, et al., "Double-Sided, Double-Type-Column 3D detectors at FBK: Design, Fabrication and Technology Evaluation", *IEEE Trans. Nucl. Sci.*, vol.55(5), pp. 2275-2284, Oct. 2008.
[6] G. Pellegrini, M. Lozano, M. Ullan, R. Bates, C. Fleta, D. Pennicard, "First double-sided 3-D detectors fabricated at CNM-IMB", *Nuclear Instrum. Methods A*, vol. 592, pp. 38-43, 2008.
[7] A. Zoboli, et al., "Functional Characterization of 3D-DDTC Detectors Fabricated at FBK-irst", *2008 IEEE Nuclear Science Symposium*, Dresden (Germany), Oct. 19-25, 2008, Conference Record, Paper N34-4.
[8] A. Zoboli et al., "Initial results from 3D-DDTC detectors on p-type substrates", to be published in *Nuclear Instrum. Methods A* (doi:10.1016/j.nima.2009.08.010).
[9] G.-F. Dalla Betta, et al., "Development of 3D-DDTC pixel detectors for the ATLAS upgrade", presented at 7th HSTD, Hiroshima (Japan), Aug. 29 – Sept. 1 2009, submitted for publication to *Nuclear. Instrum. Methods A*.
[10] A. Zoboli et al., "Characterization and modeling of signal dynamics in 3D-DDTC detectors", to be published in *Nuclear Instrum. Methods A* (doi:10.1016/j.nima.2009.09.035).
[11] F. Campabadal et. al., "Design and performance of the ABCD3TA ASIC for readout of silicon strip detectors in the ATLAS semiconductor tracker", *Nuclear Instrum. Methods A*, vol. 552, pp. 292-328, 2005.
[12] G.-F. Dalla Betta, et al., "Performance evaluation of 3D-DDTC detectors on p-type substrates", presented at 11th ESSD, Wildbad Kreuth (Germany), June 7-11, 2009, submitted to *Nuclear. Instrum. Methods A*.
[13] I. Perić, et al., "The FEI3 readout chip for the ATLAS pixel detector", *Nuclear. Instrum. Methods A* 565, pp. 178-187, 2006.
[14] SELEX Sistemi Integrati, Roma, Italy. http://www.selex-si.com
[15] http://physik2.uni-goettingen.de/~jgrosse/TurboDAQ.
[16] A. Andreazza, et al., "ATLAS Pixel Module Electrical Tests", ATLAS Project Document No: ATL-IP-QP-0144.
[17] G. Aad, et al., "ATLAS pixel detector electronics and sensors", *JINST*, vol. 3, P07007, 2008.